\documentclass{sig-alternate}
\usepackage{mathptmx} 

\usepackage{fancyhdr}
\usepackage[normalem]{ulem}
\usepackage[hyphens]{url}
\usepackage[sort,nocompress]{cite}
\usepackage[final]{microtype}
\usepackage[keeplastbox]{flushend}
\usepackage{subfigure}

\usepackage{listings}
\usepackage{multirow}
\usepackage{amsmath,amssymb,amsfonts}
\usepackage{xcolor}

\usepackage[bookmarks=true,breaklinks=true,letterpaper=true,colorlinks,linkcolor=black,citecolor=blue,urlcolor=black]{hyperref}

\newcommand{\sectiondistance}{-0.4cm}
\newcommand{\papertitleabbr}{PMUSpill}
\newcommand{\papertitle}{The Counters in Performance Monitor Unit that Leak SGX-Protected Secrets}

\title{\papertitleabbr: \papertitle} 

\author{
Pengfei Qiu, Yongqiang Lyu, Haixia Wang, Dongsheng Wang, Chang Liu\\
       \affaddr{Tsinghua University, Beijing, China.}\\
       \email{qpf15@tsinghua.org.cn, \{luyq, hx-wang, wds\}@tsinghua.edu.cn, cliu21@mails.tsinghua.edu.cn}
\and
Qiang Gao, Chunlu Wang\\
       \affaddr{Beijing University of Posts and Telecommunications, Beijing, China.}\\
       \email{\{2020140851, wangcl\}@bupt.edu.cn}
\and 
Rihui Sun\\
       \affaddr{Harbin Institute of Technology, Harbin, China.}\\
       \email{srh@stu.hit.edu.cn}
\and
Gang Qu\\
       \affaddr{University of Maryland, College Park, USA.}\\
       \email{gangqu@umd.edu}
}

\begin{document}
\maketitle
\pagestyle{plain}


\begin{abstract}
Performance Monitor Unit (PMU) is a significant hardware module on the current processors, which counts the events launched by processor into a set of PMU counters. Ideally, the events triggered by instructions that are executed but the results are not successfully committed (transient execution) should not be recorded. However, in this study, we discover that some PMU events triggered by the transient execution instructions will actually be recorded by PMU. Based on this, we propose the \papertitleabbr\ attack, which enables attackers to maliciously leak the secret data that are loaded during transient executions. The biggest challenge is how to encode the secret data into PMU events. We construct an instruction gadget to solve this challenge, whose execution path that can be identified by PMU counters represents what values the secret data are. We successfully implement the \papertitleabbr\ attack to leak the secret data stored in Intel Software Guard Extensions (SGX) (a Trusted Execution Environment (TEE) in the Intel's processors) through real experiments. Besides, we locate the vulnerable PMU counters and their trigger instructions by iterating all the valid PMU counters and instructions. The experiment results demonstrate that there are up to 20 PMU counters available to implement the \papertitleabbr\ attack. We also provide some possible hardware and software-based countermeasures for addressing the \papertitleabbr\ attack, which can be utilized to enhance the security of processors in future. 
\end{abstract}

\section{Introduction}
\vspace{\sectiondistance}
Performance Monitor Unit (PMU)~\cite{Intel2015Intel, liu2016catalyst, cho2020real} is a widely enabled hardware module on processors nowadays including those by Intel, ARM, and AMD, which contains a set of counters to count the events triggered during run-time of the core and its memory system such as cache hit/miss, write-back transaction, streaming store request, machine clear, etc. It is essentially designed as a tool to collect the activities of processor for software developers to analyze and optimize their codes. However, PMU provides plentiful information about the behaviors of processor, which provides opportunities for researchers to reverse engineering the design details of black-box processors~\cite{schwarz2019zombieload, van2019ridl, ragab2021crosstalk, mambretti2019speculator} or speculate the keys of encryption algorithms such as Advanced Encryption Standard (AES)~\cite{gotzfried2017cache}, Rivest-Shamir-Adleman (RSA)~\cite{bhattacharya2015watches}, and Elliptic Curves Cryptography (ECC)~\cite{asvija2020template}. While this study shows a new security issue caused by PMU that is not studied before as we know: leaking the secret data that are loaded during transient executions.

Processor designers have integrated several high-performance technologies (e.g. out-of-order execution, speculative execution, Micro-architectural Data Sampling (MDS)~\cite{Intel2015Intel}, et al.) into current processors to enhance the instruction throughput. Those technologies enable processors to perform later instructions even when previous instructions have not retired. Such execution of “invisible” instructions is referred to the transient execution, whose execution results will be roll backed if an exception handling delay on out-of-order execution~\cite{lipp2018meltdown, van2018foreshadow}, a misprediction on speculation~\cite{kocher2019spectre}, or a microcode-assist event on MDS~\cite{schwarz2019zombieload, canella2019fallout} happens. However, in this study, we discover that some events happened in the transient executions will also be recorded by PMU counters. In other words, the processor will not roll back those PMU counters regardless of the executed instructions retire or not. This is not studied by the prior PMU-based attacks as their attacks are conducted by measuring the events triggered by retired instructions. Based on the discovery, we propose the \papertitleabbr\ attack, a new kind of PMU-based side channel attack that is available to achieve transient execution attacks. In \papertitleabbr\ attack, the secret data are encoded into the events that will be recorded by PMU during the transient executions, which are recovered by monitoring whether the corresponding PMU counters are changed after the transient executions. 

Finding a way to encode the secret data into PMU events during transient executions is the biggest challenge when implementing \papertitleabbr\ attack. Directly using the secret data as instruction operands is not a good choice, this is because that PMU events are usually determined by the instruction opcodes instead of operands. In this study, we design an instruction gadget to fully solve this challenge. The instruction gadget contains a transient execution in which a controllable variable is compared to the secret data. The execution path when the variable equals to the secret data is different from that when they are unequal. 
Besides, the instructions of the two execution paths are carefully selected so that the events triggered by them are different. As a result, the events triggered by the execution of the instruction gadget reflect which execution path is taken and therefore reveal whether the controllable variable and secret data are the same to attackers.

In order to locate the vulnerable PMU counters that are available to implement \papertitleabbr\ attack and which instructions can trigger their corresponding events (called trigger instructions in this study), we feed the two execution paths of the instruction gadget with the instruction iterated through the uops.info data set~\cite{abel2019uops} (an instruction data set for the Intel X86 architecture) and test all the valid PMU counters after the execution of the instruction gadget. The PMU counters whose increments when the variable equals to the secret data are different from that when they are diverse are vulnerable counters and the corresponding instructions are the trigger instructions. With our investigation, there are up to 20 vulnerable PMU counters in our experiment devices whose processor is Intel$^\circledR$ Core$^\text{TM}$ i7-6700 (the total number of valid PMU counters is 217). We successfully utilize them to leak the secret data protected by the Intel Software Guard Extensions (SGX) (a Trusted Execution Environment (TEE) in the Intel's processors), which illustrates that the \papertitleabbr\ attack is a real threaten to the computer system. Besides, our experiment results demonstrate that the throughput of \papertitleabbr\ attack can up to 575.3 byte per second (Bps) with an error rate of less than 2\% on average.

Disabling the vulnerable PMU counters is an efficient solution to defend against \papertitleabbr\ attack, however, this will take ineluctable side effects that software developers cannot use those PMU counters to analyze the performance of their codes. Moreover, even we test all the valid PMU counters in our experiment devices and all the instructions in the uops.info data set~\cite{abel2019uops}, we cannot guarantee that we find out all of the vulnerable PMU counters especially the PMU counters that are private for some processors. \papertitleabbr\ attack is implemented by monitoring the events happened in the transient executions, which proves that it is not a good design to expose the transient activities of the processor to the users even the privileged users. Besides, the attack targets of the \papertitleabbr\ attack are the procedures that are performed in the TEE. Therefore, in this work, we propose two methods to mitigate \papertitleabbr\ attack: 1) renaming the PMU counters when transient executions are performing and only updating the real PMU counters when transient instructions successfully retire. This is a hardware-based solution and cannot provide protections to the existing devices. 2) similar to the methods that are designed by Intel to mitigate the vulnerability CVE-2019-11157~\cite{qiu2019voltjockeySGX, qiu2020voltjockey}, ensuring that the PMU and SGX cannot be simultaneously enabled by the microcode update is a useful software-based countermeasure. This method can be deployed on the existing systems.

In sum, we make the following contributions in this work.
\begin{itemize}
\item We discover that some events triggered by transient executions will also be recorded by PMU, which will unintentionally leak the behaviors of the transient executions and cause secret data leakages. We therefore propose \papertitleabbr\ attack, a new kind of side channel attack that is able to infer the transiently loaded secret data with the help of PMU.
\item We design an instruction gadget to encode secret data into transient execution events, in which a controllable variable is compared to the secret data and the execution path after the comparison that can be identified by PMU counters represents which the secret data are. 
\item We go through all the PMU counters in Intel$^\circledR$ Core$^\text{TM}$ i7-6700 processor with all valid instructions, and find that there are up to 20 vulnerable PMU counters available to implement \papertitleabbr\ attack. The applications of \papertitleabbr\ attack is verified by obtaining the Intel SGX-protected secret data through real experiments. 
\end{itemize}

\section{Preliminaries and Related Work}
\vspace{\sectiondistance}

\subsection{Side Channel Attack}
\vspace{\sectiondistance}
Side channel attack is a kind of widely studied attack method in the field of hardware security, which leaks secret data from a victim process to the attacker process by monitoring the information (e.g. instruction execution time, power consumption, electromagnetic, branch prediction, et al.) from the implementation of hardware system rather than using software vulnerabilities in the implemented algorithms. The monitored information is usually unintentionally leaked by the hardware, which results that it is difficult to address the side channel attack. Side channel attack have been widely used to infer the keys of encryption algorithms that are not easy to break with brute force attacks such as Data Encryption Standard (DES)~\cite{hu2019effective}, AES~\cite{kayaalp2016high, jadid2021survey}, RSA~\cite{lipp2021platypus, gras2020absynthe, gras2018translation}, EdDSA~\cite{weissbart2019one, gras2018translation}, and ECC~\cite{alam2021nonce}. 

There are several types of side channel attacks proposed such as cache side channel attack~\cite{liu2015last}, timing side channel attack~\cite{hund2013practical}, power side channel attack~\cite{wei2018know}, electromagnetic side channel attack~\cite{yilmaz2019electromagnetic}, acoustic side channel attack~\cite{deepa2013overview}, branch prediction side channel attack~\cite{evtyushkin2018branchscope}, et al. The cache side channel attack is the most investigated attack method, which is implemented based on the fact that the execution time for a data load operation that misses the cache is obviously higher than that when hits the cache. Prime+Probe~\cite{tromer2010efficient}, Evict+Time~\cite{osvik2006cache, van2021cacheout}, Flush+Flush~\cite{gruss2016flush+}, and Flush+Reload~\cite{yarom2014flush+} are the four famous cache side channel attacks. Among those attacks, Flush+Reload attack~\cite{yarom2014flush+} is very easy to implement and widely utilized in recovering the secret data obtained by transient execution vulnerabilities. 

Randomization, partition, and sensitive data differentiating technologies have become momentous countermeasures against side channel attacks \cite{gundu2014memory} by randomizing the data access, isolating the attacker and victim processes, and allowing fine-grained control of the secret data. Besides, some secure technologies have also been proposed to address side channel attacks such as kernel separation \cite{buerki2013muen}, virtualization \cite{godfrey2014preventing}, and trusted computing like Intel Trusted Execution Technology (TXT) \cite{futral2013intel}, SGX \cite{costan2016intel}, and ARM TrustZone \cite{arm2009security}.

\subsection{Performance Monitor Unit (PMU)}
\vspace{\sectiondistance}
PMU is a widely-enabled on-chip hardware that measures the processor's various architectural and micro-architectural events into several special registers (PMU counters). With PMU, we can gain insight into the processor's behaviors such as the number of instructions that are assigned to a particular port, the number of cache misses/hits at different levels in the memory hierarchy, the number of mis-predicted branches, et al. It is initially designed for software developers to analyze the code execution efficiency and therefore help them to optimize their applications. However, because it is capable to disclose the impacts of an instruction on the processor in a fine-grained manner, the researchers have also try to utilize the PMU to achieve other aims such as reverse engineering the black-box processors~\cite{van2019ridl, schwarz2019zombieload, ragab2021crosstalk}, and inferring the keys of encryption algorithms~\cite{gotzfried2017cache, bhattacharya2015watches, asvija2020template}. 

{\it Reverse engineering the black-box processors.} Most of the commercial processors nowadays are black-box processors, it is very difficult for researchers to analyze their security with the limited public information. The PMU provides an opportunity for researchers to pry into the inner designs of the black-box processors for vulnerability analysis and mining. For example, Van et al. and Schwarz et al. find the RIDL~\cite{van2019ridl} and ZombieLoad~\cite{schwarz2019zombieload} vulnerabilities, respectively, but how to find the leakage source become a challenge. In their work, they successfully locate the leakage source using PMU; Ragab et al. ~\cite{ragab2021crosstalk} find a micro-architectural unit (named staging buffer) that is not published in the official documents by reverse engineering the Intel processors. They then verify that the staging buffer can cause secret data leakage (the vulnerability's name is CrossTalk); Ragab et al. use the machine clear-related PMU counters to recover the processor's behaviors when machine clear events happen~\cite{ragab2021rage}, they find that some mechanisms that will trigger the machine clear are unsafe (the attacks are called Floating Point Value Injection (FPVI) and Speculative Code Store Bypass (SCSB)); Mambretti et al.~\cite{mambretti2019speculator} utilize PMU to reverse engineering the processor's activities in order to build a deeper understanding of speculative executions.

{\it Inferring the keys of encryption algorithms.} PMU provides a lot of information about the execution processes of instructions, which have been verified useful to achieve side channel attacks for inferring the keys of encryption algorithms. For example, Uhsadel et al.~\cite{uhsadel2008exploiting} and G\"{o}tzfried et al.~\cite{gotzfried2017cache} use the PMU counters that measure the number of L1-cache hits or misses to implement the access-driven cache-timing attack on the AES implementation in the OpenSSL library; Bhattacharya et al.~\cite{bhattacharya2015watches} propose to use the PMU counters that measure the number of branch misses to compromise keys of RSA; Asvija et al.~\cite{asvija2020template} propose to attack the ECC by using the power-related counters to implement the template attack. 

However, the current PMU-based side channel attacks only utilize some special PMU counters to speculate the keys of encryption algorithms. Besides, they measure the PMU events for retired instructions to achieve the attacks and do not study whether PMU can record the events activated by instructions that does not successfully retire. Moreover, they do not study whether the PMU counters can be utilized in other malicious attacks such as implementing the transient execution attacks. In this study, we find that PMU has the ability to measure the events happened during transient executions and therefore can be treated as the side channel signals to leak secret data processed by the transient executions, which is a new kind of side channel attack and is innovative compared to the existing PMU-based side channel attacks. 

\subsection{Transient Execution Vulnerabilities}
\vspace{\sectiondistance}
\label{Transientvu}
Modern processors have introduced several technologies to optimize the instruction pipeline and data access for gaining a high performance such as speculative execution, out-of-order execution, and Micro-architecture Data Sampling (MDS), et al.~\cite{Intel2015Intel}. Although the instructions may not be executed sequentially, the reorder buffer~\cite{rosiere2012out} ensures that the instructions are committed in order. Because of the processor's exception delay processing mechanism, if a misprediction on speculation~\cite{kocher2019spectre}, an exception on out-of-order execution~\cite{lipp2018meltdown, van2018foreshadow}, or a microcode-assisted event on MDS~\cite{schwarz2019zombieload, canella2019fallout} happens during the execution of an instruction, the subsequent instructions may be executed but the results may not be submitted, and the processor will roll back to eliminate the instruction's impacts on the processor's architecture layer. These instructions are called transient instructions, and the execution process of the transient instructions is called transient execution.

However, even the transient executions do not affect the processor's architecture layer, they may affect the processor's micro-architecture layer, which leaves a set of transient execution vulnerabilities that can be deliberately utilized by attackers to maliciously leak the secret data. For example, the attackers can utilize the transient instructions to change the layouts of processor's micro-architecture modules such as cache~\cite{lipp2018meltdown}, Translation Look-aside Buffer (TLB)~\cite{gras2018translation}, branch predictor~\cite{kocher2019spectre, evtyushkin2018branchscope}, etc. based on the secret data, and then recover the secret data using side channel attacks. 

Currently, the transient execution attacks can be classified into three types according to the technologies that cause the vulnerabilities: out-of-order execution-based attacks, speculative execution-based attacks, and MDS-based attacks. In the first type, the attacker triggers an exception and then encodes the secret data into the processor's micro-architecture layer during the following transient execution instructions enabled by the out-of-order execution mechanism. The represented attacks of this type include Meltdown~\cite{lipp2018meltdown}, Foreshadow~\cite{van2018foreshadow}, and so on. In the second type, the attacker first induces some speculative executions to poison the branch predictor and then changes the layouts of the processor's micro-architecture layer by causing a misprediction. The represented attacks of this type include Spcetre V1~\cite{kocher2019spectre}, Spcetre V2~\cite{hill2019spectre}, and Spectre RSB~\cite{koruyeh2018spectre}. In the third type, the attacker first triggers a microcode-assist data access event and then put the secret data-based information into the processor's micro-architecture layer. The represented attacks of this type include Zombieload\cite{schwarz2019zombieload}, RIDL~\cite{van2019ridl}, Fallout~\cite{canella2019fallout}, and Medusa~\cite{moghimi2020medusa}. The leakage sources of this type of attacks are Line Fill Buffers (LFB)~\cite{schwarz2019zombieload, van2019ridl}, Store Buffer (SB)~\cite{canella2019fallout}, and Read Buffer (RB), et al. Besides, this type of exploits sample the required data from some in-flight data instead of specifying a full physical or virtual address.

\subsection{Trusted Execution Environment}
\vspace{\sectiondistance}
\label{TEEBackground}
TEE~\cite{sabt2015trusted} is a technology that helps manufacturers, service providers, and consumers protect their devices and sensitive data from disclosure or modification by providing a secure isolated execution environment. The Trusted Computing Base (TCB) of the TEE only includes the processor package, therefore, its private codes and data cannot be viewed or modified by the malicious codes outside the TEE even the codes has a full privileges on the same processor. The most famous TEEs are ARM TrustZone~\cite{arm2009security} and Intel SGX~\cite{costan2016intel}. The attack targets of this study are the data owned by Intel SGX. Therefore, we mainly overview Intel SGX in this section. 

Intel SGX is a set of instruction set extensions for Intel processors that provide security of integrity and confidentiality for trusted codes executed on a malicious machine~\cite{costan2016intel, mckeen2016intel}. The trusted codes are executed in the enclave of SGX. SGX has a strict isolation mechanism for the physical memory of the enclave to ensure that the data and codes of the enclave cannot be monitored and modified by applications or advanced system software in an untrusted environment (the rich execution environment), such as Operating System (OS), hypervisor, Basic Input Output System (BIOS), System Management Mode (SMM), etc.

SGX and the rich execution environment~\cite{jang2015secret} share a large amount of processor resources, such as Cache, TLB, LFB, etc., which provide attackers with the possibility of attacks. Over the past few years, researchers have found that SGX's security still needs to be improved. It has always been challenged by side channel attacks~\cite{gotzfried2017cache, brasser2017software, bhattacharya2015watches, asvija2020template}, but transient execution attacks have also been found shortcomings of Intel SGX's security in recent years. For example, the Spectre-like attack proposed by Chen et al.~\cite{chen2019sgxpectre} can poison the Branch Target Buffer (BTB) in the enclave and uses the processor's branch prediction mechanism to leak trusted data; ZombieLoad attack~\cite{schwarz2019zombieload} uses the LFB to leak the SGX-protected data; Foreshadow attack~\cite{van2018foreshadow} maps the physical address of the enclave to an inaccessible virtual address, whose access will generate a faulting load. Taking advantage of the processor's out-of-order execution mechanism, Foreshadow can recover the enclave's secret data.

\section{Overview of \papertitleabbr\ Attack}
\vspace{\sectiondistance}
As mentioned in the Section~\ref{Transientvu}, the transient execution attacks are a combination of transient execution vulnerabilities and side channel attacks. The transient execution vulnerabilities put the secret data into the processor's micro-architecture layer, and side channel attacks leak the secret data to the processor architecture layer. Therefore, the side channel attack is also a key link of transient execution attacks, and the availability of side channel attacks will also directly affect the availability of transient execution attacks. In this study, we propose the \papertitleabbr\ attack, which is a new kind of side channel attack that is capable of bringing the secret data obtained by transient execution vulnerabilities to attackers.

In this section, we first present the motivation and threat model of this work. Next, we demonstrate the attack targets of \papertitleabbr\ attack and how it works. Finally, we show the challenges that should be addressed when implementing the attack.

\subsection{Motivation}
\vspace{\sectiondistance}
Generally, the executed results of transient executions will be abandoned and the statuses of some processor's hardware modules (e.g. registers and pipeline) will be roll backed to ensure the correctness of the processor. However, in this study, we discover that some events triggered in the transient executions will be recorded by PMU counters whose values are not be roll backed as other processor modules, which will unintentionally leverage the interaction between the transient execution instructions and hardware activities. This discovery gives the attackers chances to recover the secret data by monitoring the behaviours of the transient execution instructions through PMU counters. Two nature questions are ``how to utilize such a finding to leak the data processed by transient executions?'' and ``how serious of the data leakage?''. In this work, we propose the \papertitleabbr\ attack to answer the two questions, which is a new kind of side channel attack method that builds a bridge between attackers and the secret data that is acquired in transient executions. 

\subsection{Assumption and Threat Model}
\vspace{\sectiondistance}
In the \papertitleabbr\ attack, we have three assumptions:
\begin{itemize}
\item PMU is enabled by the target processor and some of them can record the events occurred during transient executions. As far as we know, all of the modern Intel, ARM, and AMD processors provide PMU as a performance profiling technology for software developers. This work demonstrates that several PMU counters in Intel processors can count the events triggered by transient executions. Besides, we have verified that some PMU counters in ARM processors are also able to monitor the events triggered by the transient executions. Therefore, this assumption is true for a range of the current processors.
\item The secret data can be acquired in the transient executions. This has been verified by a lot of reported transient execution attacks such as Meltdown~\cite{lipp2018meltdown}, Spectre~\cite{kocher2019spectre}, Foreshadow~\cite{van2018foreshadow}, RIDL~\cite{van2019ridl}, Fallout~\cite{canella2019fallout}, ZombieLoad~\cite{schwarz2019zombieload}, Medusa~\cite{moghimi2020medusa}, CrossTalk~\cite{ragab2021crosstalk}, FPVI~\cite{ragab2021rage}, and SCSB~\cite{ragab2021rage} et al. Although those attacks have been addressed by some countermeasures including software updates or hardware redesign, we can not guarantee that our processors do not have the undisclosed transient execution vulnerabilities. The new transient execution attacks are continually reported in the recent years is an evidence to prove that we may not discover all of the transient execution vulnerabilities.
\item The attackers have a root privilege. This assumption gives the attackers ability to read the PMU counters. As presented in the Section \ref{TEEBackground}, the root privilege is out of the TCB of TEE. Therefore, assuming attackers have the root privilege is permissible when the victim procedures are in TEE. Besides, there are many legal or illegal ways for attackers to gain such privilege~\cite{niu2014overview, xiao2016one}.
\end{itemize}

Under such assumptions, we make the threat model of this study as: the attackers try to recover the secret data that is acquired in the transient executions by measuring the PMU events and therefore achieve the transient execution attacks.

\subsection{Attack Targets}
\vspace{\sectiondistance}
\papertitleabbr\ attack is a side channel attack, whose ability is to recover the secret data accessed in transient executions. Because that reading PMU counters requires a root privilege and this does not violate the TCB of TEE, the secret data stored in TEE are the appropriate attack targets of \papertitleabbr\ attack. The processor utilized in our experiment device is an Intel processor, whose TEE is SGX. As presented in the Section \ref{TEEBackground}, although SGX enables a lot of security mechanisms to guarantee that its private data cannot be attacked by a lot of software-based vulnerabilities~\cite{costan2016intel}, it cannot prevent the data leakage from the processor hardware vulnerabilities such as transient execution vulnerabilities. In fact, several transient execution attacks have been verified feasible to attack the SGX-private data including Foreshadow~\cite{van2018foreshadow}, ZombieLoad~\cite{schwarz2019zombieload}, and Spectre~\cite{chen2019sgxpectre}. In this study, we utilize \papertitleabbr\ attack to implement the Foreshadow attack through real attacks, which demonstrates the efficiency of \papertitleabbr\ on leaking the secret data of SGX.

\subsection{Attack Steps of \papertitleabbr\ Attack}
\begin{figure}[htb]
\begin{center}
\includegraphics[width=0.98\columnwidth]{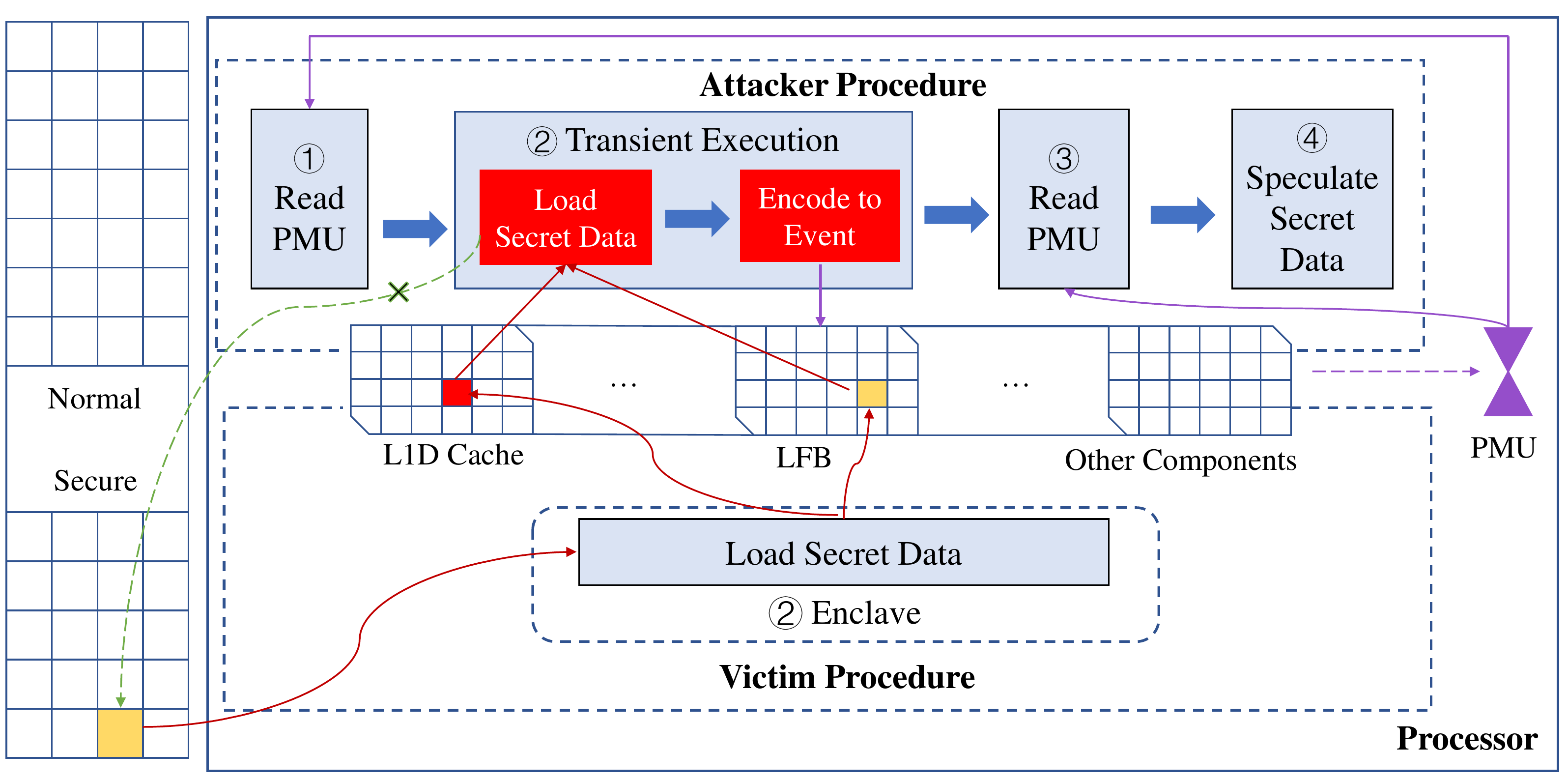}
\caption{The overview and attack steps of \papertitleabbr\ attack. \textcircled{1} getting the initial value of the PMU counter; \textcircled{2} acquiring the secret data and encoding them into the PMU event using the transient execution instructions; \textcircled{3} re-accessing the PMU counter to get its new value; \textcircled{4} inferring the secret data based on the change of the PMU counter's value.}
\label{fig:overall}
\end{center}
\end{figure}

\vspace{\sectiondistance}
Figure \ref{fig:overall} depicts the overview and attack steps of \papertitleabbr\ attack, where PMU plays a role to expose the secret data that is maliciously acquired in transient executions to attackers. The attacker procedure is performed in the rich execution environment. It cannot directly read the secret data stored in TEE but can access them transiently through transient executions. There are four steps in the attacker procedure to achieve the \papertitleabbr\ attack: 1) reading the PMU counter to get its initial value; 2) accessing the secret data in the transient executions and encoding them into the PMU event by executing some special instructions (trigger instructions) based on the secret data; 3) reading the PMU counter to get its new value; 4) inferring the secret data based on whether the PMU counter is changed or its change ranges. What should be mentioned is that all the PMU counters that are updated in the transient executions may be available to implement the \papertitleabbr\ attack.

\subsection{Challenges}
\vspace{\sectiondistance}

\subsubsection{Encoding the Secret Data into PMU Events}
\vspace{\sectiondistance}
This is the biggest challenge that should be addressed. Generally, it is not a good choice to encode the secret data into PMU events by directly treating them as the instruction operands (operation data of the instruction) because that the triggered PMU events are mainly decided by the instruction opcodes instead of operands. Of course, the instruction operands may also have influences on the PMU events that are triggered such as the data access address, the target address of the jump instructions, or the operation data of some vector operation instructions. However, those PMU events are usually triggered by some special operation data and therefore requires to carefully analyze the functions of every instruction to decide how to manipulate the operation data. For example, the operation data of a memory-access instruction may influence the event of L1 cache miss, but this event only happens when the data referenced by the operation data of the instruction are not in the L1 cache. Besides, directly treating the secret data as the instruction operands has little influence on this study as our aim is to verify the efficiency of the \papertitleabbr\ attack and find out the vulnerable PMU counters as well as their possible trigger instructions. 

In this study, we design an instruction gadget to solve this challenge, where the secret data are encoded into the execution path (can be measured by the PMU counters) of the instruction gadget. The details of the instruction gadget will be described in the Section \ref{encodesecret},

\subsubsection{Searching for the Vulnerable PMU Events}
\vspace{\sectiondistance}
Accessing the secret data in the attacker procedure will trigger an exception because that the attacker procedure has no privilege to access them, therefore, the secret data is transiently owned by the attacker procedure. Different from the retired executions whose behaviours will be naturally recorded by the corresponding PMU counters, not all of the PMU events will be recorded by their PMU counters in the transient executions. In other words, we cannot encode the secret data into the PMU events that are only updated when the instructions successfully retire. 

In this study, we give the different execution paths of the instruction gadget different instructions (those instructions are selected from the uops.info data set~\cite{abel2019uops}) and measure all of the valid PMU counters before and after the instruction gadget is executed. The PMU counters whose values are prospectively changed are the available PMU counters to implement \papertitleabbr\ attack and the corresponding instructions are the trigger instructions. Section \ref{EventsIteration} and Section \ref{VulnerablePMU} shows our experiment settings and results about the vulnerable PMU counters and their trigger instructions, respectively.

\subsubsection{Reducing the Noises}
\vspace{\sectiondistance}
PMU is shared by all of the procedures executed on the same core, therefore, there may be some noises from other procedures that are scheduled onto the attacker core (the core that performs the attacker procedure) by the OS. In this study, we mitigate those noises by setting the attacker core as an isolated core. Through this, the OS cannot schedule other procedures onto the attacker core, but we can still bind the attacker procedure to the attacker core. This does not violate our assumption as owning the root privilege is enough to achieve this setting. In our experiments, when we do not isolate the attacker core, we get an error rate of 5.17\% with the same attack parameters described in the Section \ref{Experimentresults}.

Using the software codes to capture and handle the exceptions caused by the operation of accessing the secret data in the attacker procedure will also take some noises because that the exception handling codes may also trigger some PMU events. Therefore, it is a better choice to utilize the Intel Transactional Synchronization Extensions (TSX)~\cite{Intel2015Intel, lipp2018meltdown, schwarz2019zombieload, van2018foreshadow} to suppress those noises. The Intel TSX is a hardware feature of the Intel processors who treats the executed instructions as an atomic operation and will automatically handle the exceptions. TSX has been widely utilized to handle the exceptions in the existing transient execution attacks including Meltdown~\cite{lipp2018meltdown}, ZombieLoad~\cite{schwarz2019zombieload}, Foreshadow~\cite{van2018foreshadow}, et al. According to our investigations, the error rate when using software codes to suppress the exceptions is also acceptable (the experiment results are shown in the Table \ref{table:results}) because that the events triggered by the exception handling codes are certainly fixed. Therefore, for the target processors that do not enable the TSX, we can still achieve an efficient \papertitleabbr\ attack. 

\section{Encoding Secret Data into PMU Events in Transient Executions}
\vspace{\sectiondistance}
\label{encodesecret}
In this study, we construct an instruction gadget to encode the secret data into PMU events in transient executions, which is presented in the Listing~\ref{Instructiongadget} (written in the C language). 

\lstset{
    basicstyle=\footnotesize\ttfamily,
    keywordstyle=\color{purple}\bfseries,
    identifierstyle=\color{brown!80!black},
    commentstyle=\color{gray},
    showstringspaces=false,
    framexleftmargin=6mm,
    frame = box,
}
\begin{lstlisting}[language=C,frame=trbl,captionpos=b, frameround = tttt,numbers = left,xleftmargin=2.5em,xrightmargin=1.2em,caption = The constructed instruction gadget, label=Instructiongadget,]
char temp = `a', address[6], value[6]
for(m = 0; m < 5; m++){
  address[m] = &temp;
  value[m] = `b';
}
address[5] = secret_data_address;
for(j = 0; j < 256; j++){
  value[5] = j;
  for(m = 0; m < 6; m++){
    pmu_start = read_pmu(pmu_number);
    pmu_end = read_pmu(pmu_number);
  }
  value[5] = j;
  for(m = 0; m < 6; m++){
    pmu_start = read_pmu(pmu_number);
    pmu_end = read_pmu(pmu_number);
  }
  value[5] = j;
  for(m = 0; m < 6; m++){
    pmu_start = read_pmu(pmu_number);
    pmu_end = read_pmu(pmu_number);
  }
  value[5] = j;
  for(m = 0; m < 6; m++){
    pmu_start = read_pmu(pmu_number);
    pmu_end = read_pmu(pmu_number);
  }
  for(m = 0; m < 6; m++){
    pmu_start = read_pmu(pmu_number);
    pmu_end = read_pmu(pmu_number);
  }
}
\end{lstlisting}

\subsection{The Core Instructions}
\vspace{\sectiondistance}
The instructions at line 12-28 are the core instructions of the instruction gadget. The other instructions play a role to help readers understand how to apply the \papertitleabbr\ attack to recover the secret data in the transient execution attacks. The instruction at line 14 loads secret data into the \texttt{rbx} register, which will trigger an exception as the attacker procedure does not have the privilege to directly load the secret data. Therefore, instructions at line 15-22 are executed but the execution results are not committed (those eight instructions are transient execution instructions in the attacks). 

The instruction at line 15 compares the \texttt{rbx} register (stores the secret data whose length of bits is 8 in this example listing codes) with the \texttt{rdx} register (stores the controllable comparison variable \texttt{j} whose range is 0-255 in this example listing codes). If the \texttt{rbx} register equals to the \texttt{rdx} register, the instruction at line 20 (ins2, we utilize the \texttt{mov (rcx), rax} as an example in the Listing \ref{Instructiongadget}) will be executed, otherwise, the instruction at line 17 (ins1, we utilize the \texttt{mov (rax), rax} as an example in the Listing \ref{Instructiongadget}) will be executed. The instruction at line 18 ensures that the \texttt{ins1} will not be executed if the \texttt{rbx} register does not euqal to the \texttt{rdx} register. 

As explained before, this instruction gadget can ensure that the execution path when the secret data equals to the comparison variable (the instruction that is executed is \texttt{ins1}) is different from the execution path when they are different values (the instruction that is executed is \texttt{ins2}) . If the \texttt{ins1} and \texttt{ins2} trigger different PMU events, we can judge whether \texttt{ins1} is executed or the \texttt{ins2} is executed by measuring the PMU events from their corresponding PMU counters. If we find that the \texttt{ins1} is executed, we can make a conclusion that the comparison variable equals to the secret data.

\subsection{Dealing with the Speculative Execution}
\vspace{\sectiondistance}
Because of the processor's speculative execution mechanism, if the branch predictor for the jump instruction at line 16 are not prepensely trained, the \texttt{ins2} may be speculatively executed even when the \texttt{rbx} register and the \texttt{rdx} register are unequal. This may disturb the precision of the recognition for the transient execution path. 

In order to avoid this situation, the instruction gadget firstly executes the core instructions for 5 times with the condition of the \texttt{rbx} register does not equal to the \texttt{rdx} register to train the branch predictor to predict that the next condition for the comparison instruction is false. In the training process, there is no exception triggered because that the value stored in the \texttt{rbx} register are completely controlled by the attacker procedure. In the sixth execution of the core instructions, the \texttt{rbx} register will holds the secret data. and the executed instructions will be highly corresponded to the result of the comparison instruction at line 15. 

\subsection{Identifying the Memory-Related Events}
\vspace{\sectiondistance}
Identifying the memory-related events is one of the most important aims of PMU. Actually, in our experiment device (the architecture of the processor is Skylake), Intel provides about 89 PMU counters (the number of the total available PMU counters is 217) to count the memory-related events, 54 of them are the cache hit/miss-corresponded counters. Therefore, in the instruction gadget, we also provide support for identifying some memory-related events, especially the cache hit/miss-related events. 

In the instruction gadget, the \texttt{ins1} and \texttt{ins2} are designed to use different registers for the indirect addressing (in our implementations, \texttt{ins1} use the \texttt{rax} register for indirect addressing and \texttt{ins2} use the \texttt{rcx} register for indirect addressing). Before executing the core instructions, the value referred by the \texttt{rax} register is flushed from the cache and the value referred by the \texttt{rcx} register is accessed so that it is cached. With this mechanism, the memory access will miss the cache (loading data from the main memory) when the secret data equals to the comparison variable, otherwise, the memory access will hit the cache (directly loading data from the cache memory). Consequently, the instruction gadget is able to recognize some of the memory-related events especially the cache hit/miss-related events that are triggered during the execution of the core instruction of the instruction gadget.

\section{Implementation and Experiment Results}
\label{Experimentresults}
\vspace{\sectiondistance}
We verify the usability of \papertitleabbr\ attack by implementing the Foreshadow attack to leak the data protected by SGX through real experiments. This section shows our experiment details and results.

\subsection{The Experiment Setup}
\vspace{\sectiondistance}
The victim device utilized in our demonstration experiments is Dell Optiplex 7040 desktop featuring an Intel$^\circledR$ Core$^\text{TM}$ Skylake quad-core i7-6700 processor with a 32 KiB, 8-way L1 data cache. Van et al.~\cite{van2018foreshadow} have verified that this processor is vulnerable to the Foreshadow attack. Besides, SGX and TSX are enabled by the processor. The operating system is Ubuntu 18.04 and the kernel version is 5.4.0-96. We refer to the public exploit codes\footnote{https://github.com/jovanbulck/sgx-step/tree/master/app/foreshadow} of the Foreshadow attack to achieve the data leakage from SGX.

The main modifications that we do to the public exploit codes of the Foreshadow attack is to replace the instructions that encode the secret data into the cache layout with our proposed instruction gadget (presented in Listing~\ref{Instructiongadget}) and the secret data recovery mechanism. Those two parts are also the most important elements in the transient execution attacks. Because PMU events are the side signals to recover the secret data instead of the cache layout in the \papertitleabbr\ attack, we should design a new secret data recovery mechanism. How to mitigate the uncontrollable noises and perturbations to the PMU counters is the main challenge that we should address when designing the secret data recovery mechanism. 

\subsection{Secret Data Recovery Mechanism}
\vspace{\sectiondistance}
The basic idea to infer the secret data is to recognize whether the expected PMU events are triggered from their related PMU counters. If a PMU event is triggered, its PMU counter will be increased. Therefore, finding an appropriate threshold for a PMU counter is a natural solution to judge whether its related PMU event is triggered. 
This method is utilized in our inchoate experiments, however, we find that it has three disadvantages: 
\begin{itemize}
\item We find that the increments of some PMU counters during the execution of an instruction are not always 1. Besides, the increments of some PMU counters are not 0 even when the executed instruction is \texttt{nop}. Therefore, it is impracticable to simply set the threshold as 0 for all the PMU counters. 
\item There are a set of PMU counters for each architecture, finding the threshold for all the PMU counters and all the architectures is a tough task.
\item Even we can make a pre-test to automatically decide the threshold by measuring the PMU counters when their events are triggered and not triggered, the difference of the execution environment will make the test results imprecise in the real attacks.
\end{itemize}

In this study, we design a simple but efficient secret data recovery mechanism based on the instruction gadget shown in the Listing~\ref{Instructiongadget}. The secret data that we recover once a time has 8 bits, therefore, the secret data have 256 possibilities. For recovering one byte of the secret data, we iterate the comparison variable from 0 to 255 to execute the instruction gadget. The PMU counter's value is stored for every iteration. The value whose occurrence number is 1 among the 256 values of the PMU counter means that the expected PMU event is triggered. We can therefore recover the secret data by looking for which measurement results in this value. The principle behind our design is that the value of the PMU counter when its event is triggered is different from that when the event is not triggered and each byte of the secret data must be a value from 0 to 255. Of course, for each byte of the secret data, executing the instruction gadget several times and inferring the secret data based on the majority voting will increase the recovery accuracy for the secret data. 

\subsection{Instructions and PMU Events Iteration}
\label{EventsIteration}
\vspace{\sectiondistance}
The \papertitleabbr\ attack depends on the PMU events that can be captured during the transient executions. Therefore, not all of the PMU events and their trigger instructions can be utilized to implement \papertitleabbr\ attack. In this study, we iterate all the valid PMU counters and instructions~\cite{abel2019uops} to search for the vulnerable PMU events and their trigger instructions that are available to achieve \papertitleabbr\ attack.

\subsubsection{Valid PMU Events on the Experiment Device}
\vspace{\sectiondistance}
In order to locate the vulnerable PMU events, we should first obtain all the valid PMU events on the target processor. Reviewing the Intel's Software Development Manual (SDM)~\cite{Intel2015Intel} is a doable method, but is time-expensive. In this study, we utilize a PMU tool\footnote{https://github.com/andikleen/pmu-tools} to output the valid PMU events on our experiment device. The output is a json file, which contains the information of every PMU event. Based on the output file, we extract the \texttt{umask} and \texttt{EventCode} (information to recognize the PMU event) of each PMU event. Through this, we get 214 valid PMU events. 

We verify that our obtained PMU events are included in the Intel SDM. However, from the Intel SDM, we find three PMU events (the first three PMU events in the Table \ref{table:results}) that are not included in the target processors but are available on other architectures also work on the target processor. This is an additional finding that the implementation of the processor does not strictly equals to the SDM. Therefore, we extend the number of the valid PMU events from 214 to 217.

\subsubsection{Valid Instructions on the Experiment Device}
\vspace{\sectiondistance}
The uops.info data set~\cite{abel2019uops} provides the information of the Intel X86 Instruction Set Architecture (ISA) in the format of XML file, which contains 14546 instructions. However, not all of the instructions can be successfully executed on our experiment devices. We first extract the assembly instructions from the data set and then convert them to the inline assembly instructions that fit the Linux system. Finally, we integrate those instructions into a file written in the C language one by one, the instructions that can be successfully compiled and executed are treated as the valid instructions in the experiment device. What should be mentioned is that the instructions that will raise exceptions will also be selected as they will be executed transiently in the real attacks. In our experiments, we get 3069 valid instructions.

\subsubsection{PMU Events and Instructions Iteration}
\vspace{\sectiondistance}
\label{iteartion}
Based on the valid PMU events and instructions obtained aforementioned, we iterate them to get the vulnerable PMU events and their trigger instructions to implement the \papertitleabbr\ attack. The \texttt{ins1} and \texttt{ins2} in the instruction gadget shown in Listing \ref{Instructiongadget} are the instructions that we should replace with the valid instructions. Therefore, a trigger instruction in this study is defined as a combination of \texttt{ins1} and \texttt{ins2}. Theoretically, for each combination of \texttt{ins1} and \texttt{ins2}, we should test all the 217 PMU events. Besides, in order to reduce the influence of the noises, we execute the instruction gadget 10 times for each PMU event and each combination of \texttt{ins1} and \texttt{ins2}. There are total $3069*3069$ combinations of the \texttt{ins1} and \texttt{ins2}, which means that test all the PMU events and all the combinations of the \texttt{ins1} and \texttt{ins2} needs to execute the instruction gadget $3069*3069*217*10\approx2.04*10^{10}$ times, which will cost a lot of time.

In order to speed up the iteration, we do not test all the combinations of \texttt{ins1} and \texttt{ins2} in this study. Instead, we test two scenarios: (1) setting \texttt{ins2} as \texttt{nop} and iterating \texttt{ins1} with the valid instructions (Scenario 1, S1); (2) setting \texttt{ins1} as \texttt{nop} and iterating \texttt{ins2} with the valid instructions (Scenario 2, S2). We call the instruction in the trigger instruction that is not \texttt{nop} as the efficient instruction. Through this, the execution times of the instruction gadget are reduced to $(3069+3069)*217*10\approx1.33*10^{7}$, which can be done in acceptable time. This will reduce the number of trigger instructions for the vulnerable PMU events but will have negligible influences in this study as our most significant goal is to find out the vulnerable PMU events as much as we can instead of obtaining all the trigger instructions. This will be discussed in the Section \ref{experimentanalysis}.

\subsection{Vulnerable PMU Events and Their Trigger Instructions}
\vspace{\sectiondistance}
\label{VulnerablePMU}
We show our experiment results about the available PMU events and the number of their corresponding trigger instructions to implement the \papertitleabbr\ attack in Table \ref{table:results}. There are up to 20 vulnerable PMU events, 8 of them do not rely on the special trigger instructions (those PMU events are activated by the shared modules of the processor instead of the private modules assigned to the executed instructions), but the others (marked by $^*$) rely on some special trigger instructions (those PMU events only can be activated by their related trigger instructions).

\begin{table*}[htb]
\centering
\linespread{1.2}
\caption{Vulnerable PMU events and the average error rate when leaking 10000 random bytes with \papertitleabbr\ attack.}
\label{table:results}
\scriptsize
\begin{tabular}{c|ccc|cc|cccc}
 \hline
 \multirow{3}{*}{Event's category} & \multirow{1}{*}{Serial} & \multirow{3}{*}{PMU event} & \multirow{3}{*}{Event description} & \multicolumn{2}{c|}{\#Trigger instr.} & \multicolumn{4}{c}{Error rate (\%)}\\
 \cline{5-10}
  & number & &  & \multirow{2}{*}{S1} & \multirow{2}{*}{S2} & \multicolumn{2}{c}{With TSX} & \multicolumn{2}{c}{Without TSX}  \\
   \cline{7-10}
 & & &  &   &  & S1 & S2 & S1 & S2 \\
 \hline 
 \multirow{4}{*}{BR\_MISP\_EXEC} 
              & \multirow{2}{*}{1} & ALL\_                & \multicolumn{1}{l|}{\multirow{2}{*}{All near executed branches (not necessarily retired)}}              & \multirow{2}{*}{3069} & \multirow{2}{*}{3069} &  \multirow{2}{*}{1.3} & \multirow{2}{*}{0.9} &  \multirow{2}{*}{3.2}&  \multirow{2}{*}{1.7}\\
             & & BRANCHES                          &               &                       &                       &  &  \\
              \cline{2-10}
              & \multirow{2}{*}{2}  & ALL\_                      & \multicolumn{1}{l|}{Speculative and retired mispredicted macro conditional}    & \multirow{2}{*}{3069} & \multirow{2}{*}{3069} &  \multirow{2}{*}{0.5} & \multirow{2}{*}{0.4}&  \multirow{2}{*}{2.4}&  \multirow{2}{*}{1.8}\\
              & & CONDITIONAL                          & \multicolumn{1}{l|}{branches}          &                       &                       &  &  \\
 \hline
 \multirow{2}{*}{BR\_INST\_EXEC} 
              & \multirow{2}{*}{3} & NONTAKEN\_                      & \multicolumn{1}{l|}{\multirow{2}{*}{Not taken macro-conditional branches}}    & \multirow{2}{*}{3069} & \multirow{2}{*}{3069} & \multirow{2}{*}{0.6} & \multirow{2}{*}{0.5}&  \multirow{2}{*}{0.9}&  \multirow{2}{*}{3.1}\\
              & & CONDITIONAL                          &           &                       &                       &  &  \\
 \hline
 \multirow{4}{*}{INT\_MISC} 
              & \multirow{2}{*}{4} & RECOVERY\_                      & \multicolumn{1}{l|}{Core cycles the allocator was stalled due to recovery}               & \multirow{2}{*}{3069} & \multirow{2}{*}{3069} & \multirow{2}{*}{1.6}  & \multirow{2}{*}{2.1}&  \multirow{2}{*}{2.8}&  \multirow{2}{*}{2.6}\\
              & & CYCLES                          & \multicolumn{1}{l|}{from earlier clear event for this thread}            &                       &                       &  &  \\
              \cline{2-10}
              & \multirow{2}{*}{5} & CLEAR\_RESTEER\_                & \multicolumn{1}{l|}{Cycles the issue-stage is waiting for front-end to fetch}               & \multirow{2}{*}{3069} & \multirow{2}{*}{3069} & \multirow{2}{*}{1.7} & \multirow{2}{*}{1.9}&  \multirow{2}{*}{7.0}&  \multirow{2}{*}{5.9}\\
              & & CYCLES                          & \multicolumn{1}{l|}{from resteered path}              &                       &                       &  &  \\

 \hline
\multirow{4}{*}{ICACHE\_64B} 
              & \multirow{2}{*}{6} & \multirow{2}{*}{IFTAG\_HIT}     & \multicolumn{1}{l|}{Instruction fetch tag lookups that hit in the instruction}                  & \multirow{2}{*}{3069} & \multirow{2}{*}{3069} & \multirow{2}{*}{1.2} & \multirow{2}{*}{1.3}&  \multirow{2}{*}{3.6}&  \multirow{2}{*}{6.3}\\
              & &                                 & \multicolumn{1}{l|}{cache (L1I)}                  &                       &                       &  &  \\
              \cline{2-10}
              & \multirow{2}{*}{7} & \multirow{2}{*}{IFTAG\_STALL}   & \multicolumn{1}{l|}{Cycles where a code fetch is stalled due to L1 instruction}                & \multirow{2}{*}{3069} & \multirow{2}{*}{3069} & \multirow{2}{*}{2.4} & \multirow{2}{*}{2.7}&  \multirow{2}{*}{8.3}&  \multirow{2}{*}{7.6}\\
              & &                                 & \multicolumn{1}{l|}{ cache tag miss}                &                       &                       &  &  \\
 \hline
 RESOURCE\_   & \multirow{2}{*}{8}  & \multirow{2}{*}{ANY}            & \multicolumn{1}{l|}{\multirow{2}{*}{Resource-related stall cycles}}      & \multirow{2}{*}{3069} & \multirow{2}{*}{3069} & \multirow{2}{*}{0.6} & \multirow{2}{*}{0.5}&  \multirow{2}{*}{3.8}&  \multirow{2}{*}{4.3}\\
STALLS        &             &                    &                                                                         &                       &                       &  &  \\
 \hline
 PARTIAL\_    & \multirow{2}{*}{9}  & \multirow{2}{*}{SCOREBOARD$^*$}     & \multicolumn{1}{l|}{Cycles where the pipeline is stalled due to serializing}               & \multirow{2}{*}{3068} & \multirow{2}{*}{3069} & \multirow{2}{*}{1} & \multirow{2}{*}{0.5}&  \multirow{2}{*}{4.9}&  \multirow{2}{*}{6.3}\\
RAT\_STALLS   &              &                   & \multicolumn{1}{l|}{operations}                       &                       &                       &  &  \\
 \hline
             & \multirow{2}{*}{10}  & STALLS\_                        & \multicolumn{1}{l|}{Execution stalls while memory subsystem has an}                       & \multirow{2}{*}{1241} & \multirow{2}{*}{0}   & \multirow{2}{*}{1.1} & \multirow{2}{*}{-}&  \multirow{2}{*}{5.2}&  \multirow{2}{*}{-}\\
            & & MEM\_ANY$^*$                        & \multicolumn{1}{l|}{outstanding load}                   &                       &                       &  &  \\
               \cline{2-10}
CYCLE\_ACTIVITY   & \multirow{2}{*}{11}      & CYCLES\_                        & \multicolumn{1}{l|}{\multirow{2}{*}{Cycles while memory subsystem has an outstanding load}}                       & \multirow{2}{*}{1199} & \multirow{2}{*}{6}   & \multirow{2}{*}{1.3} & \multirow{2}{*}{2.2}&  \multirow{2}{*}{6.5}&  \multirow{2}{*}{8.4}\\
      & & MEM\_ANY$^*$                       &                              &                       &                       &  &  \\
      \cline{2-10}
     & \multirow{2}{*}{12}   & CYCLES\_                        & \multicolumn{1}{l|}{\multirow{2}{*}{Cycles while L1 cache miss demand load is outstanding}}                  & \multirow{2}{*}{1}    & \multirow{2}{*}{0}    & \multirow{2}{*}{2.4} & \multirow{2}{*}{-}&  \multirow{2}{*}{7.6}&  \multirow{2}{*}{-}\\
     &         & L1D\_MISS$^*$                       &                                 &                       &                       &  &  \\
              
      \hline
\multirow{2}{*}{RS\_EVENTS} 
             & \multirow{2}{*}{13}   & \multirow{2}{*}{EMPTY\_CYCLES$^*$}  & \multicolumn{1}{l|}{Cycles when Reservation Station (RS) is empty for the}                & \multirow{2}{*}{1136} & \multirow{2}{*}{6}    & \multirow{2}{*}{2.1}  & \multirow{2}{*}{3.3}&  \multirow{2}{*}{5.5}&  \multirow{2}{*}{8.7}\\
             & &                                 & \multicolumn{1}{l|}{thread}                             &                       &                       &  &  \\
 \hline
\multirow{4}{*}{ITLB\_MISSES}
             & \multirow{2}{*}{14}   & \multirow{2}{*}{WALK\_PENDING$^*$}  & \multicolumn{1}{l|}{Page Miss Handler (PMH) is busy with a page walk for an} & \multirow{2}{*}{462}  & \multirow{2}{*}{444}   & \multirow{2}{*}{6.2} & \multirow{2}{*}{4.7}&  \multirow{2}{*}{9.4}&  \multirow{2}{*}{7.8}\\
             & &                                 & \multicolumn{1}{l|}{instruction fetch request}                               &                       &                       &  &  \\
              \cline{2-10}
              & \multirow{2}{*}{15}                & MISS\_CAUSES\_                  & \multicolumn{1}{l|}{Misses at all Instruction TLB (ITLB) levels that cause} & \multirow{2}{*}{42}   & \multirow{2}{*}{37}    & \multirow{2}{*}{7.3} &\multirow{2}{*}{4.9} &  \multirow{2}{*}{9.1}&  \multirow{2}{*}{8.4}\\
              & & A\_WALK$^*$                         & \multicolumn{1}{l|}{page walks}                                                                        &                       &                       &  &  \\
              
\hline
\multirow{2}{*}{ILD\_STALL} 
             & \multirow{2}{*}{16}   & \multirow{2}{*}{LCP$^*$}            & \multicolumn{1}{l|}{\multirow{2}{*}{Stalls caused by changing prefix length of the instruction}}                   & \multirow{2}{*}{0}    & \multirow{2}{*}{21}   & \multirow{2}{*}{-} & \multirow{2}{*}{0.8}&  \multirow{2}{*}{-}&  \multirow{2}{*}{2.9}\\
             & &                                 &                           &                       &                       &  &  \\
 \hline
EXE\_        & \multirow{2}{*}{17}   & EXE\_BOUND\_                    & \multicolumn{1}{l|}{\multirow{2}{*}{Cycles where no uops were executed}} & \multirow{2}{*}{16}   & \multirow{2}{*}{0}    & \multirow{2}{*}{1.7} & \multirow{2}{*}{-}&  \multirow{2}{*}{4.3}&  \multirow{2}{*}{-}\\
ACTIVITY     & & 0\_PORTS$^*$                        &                                                                         &                       &                       &  &  \\
 \hline
 \multirow{4}{*}{IDQ} 
              & \multirow{2}{*}{18}   & \multirow{2}{*}{MS\_CYCLES$^*$}     & \multicolumn{1}{l|}{Cycles when uops are being delivered to Instruction}         & \multirow{2}{*}{3}    & \multirow{2}{*}{0}   & \multirow{2}{*}{2.2} & \multirow{2}{*}{-}&  \multirow{2}{*}{6.6}&  \multirow{2}{*}{-}\\
              & &                                 & \multicolumn{1}{l|}{Decode Queue (IDQ)}              &                       &                       &  &  \\
               \cline{2-10}
             & \multirow{2}{*}{19}    & \multirow{2}{*}{MS\_MITE\_UOPS$^*$} & \multicolumn{1}{l|}{Uops initiated by MITE and delivered to IDQ while}               & \multirow{2}{*}{3}    & \multirow{2}{*}{0}    & \multirow{2}{*}{3.4} & \multirow{2}{*}{-}&  \multirow{2}{*}{5.0}&  \multirow{2}{*}{-}\\
            &  &                                 & \multicolumn{1}{l|}{Microcode Sequenser (MS) is busy}                                           &                       &                       &  &  \\

 \hline
 DTLB\_LOAD\_ & \multirow{2}{*}{20}   & MISS\_CAUSES\_                  & \multicolumn{1}{l|}{Load misses in all Data TLB (DTLB) levels that cause}                            & \multirow{2}{*}{3}    & \multirow{2}{*}{0}    & \multirow{2}{*}{0.8} & \multirow{2}{*}{-} &  \multirow{2}{*}{4.2}&  \multirow{2}{*}{-}\\
MISSES       & & A\_WALK$^*$                         &  \multicolumn{1}{l|}{page walks}                        &                       &                       &  &  \\
 \hline
\multicolumn{4}{c|}{Average}                                                                                               &  1586 &  1406 & 2.07  & 1.91 & 5.3 & 5.4\\
\hline
\end{tabular}
\end{table*}

\subsection{Throughput and Error Rate}
\vspace{\sectiondistance}
The throughput and error rate are two major factors to evaluate a side channel attack. For every scenario described in the Section \ref{iteartion}, the number of instructions that are executed is the same. Besides, the trigger instructions (\texttt{ins1} and \texttt{ins2}) are executed transiently, they have negligible effects on the throughput. Moreover, the time cost by reading the PMU counters and recovering the secret data from the values of the PMU counters is similar for different PMU events and different trigger instructions. Therefore, the throughput of \papertitleabbr\ attack is not mainly decided by the PMU events and their trigger instructions. It is mainly decided by the execution rounds of the instruction gadget for recovering one byte of the secret data and the execution time of the instruction gadget once a time. In our experiments, the instruction gadget is executed 10 rounds to recover one byte of the secret data and the throughput of the \papertitleabbr\ attack is about 575.3 byte per second (Bps) when leaking 10000 random bytes according to our experiments.

Different from the throughput, the error rate of \papertitleabbr\ attack is decided by the individual PMU events. This is because that the efforts for triggering the different PMU events are different. Some PMU events are easy to trigger but some is hard to trigger. Besides, according to our experiments, a PMU event is not always activated by their trigger instructions. 
We measure the average error rate of the \papertitleabbr\ attack on leaking the secret data who has 10000 random bytes with different PMU counters and different trigger instructions (for the PMU events whose the number of trigger instructions in every scenario is larger than 200, it will cost several time to evaluate the error rate for all of trigger instructions. Therefore, we randomly select 200 trigger instructions to calculate the average error rate). The experiment results are shown in the Table \ref{table:results}, which illustrate that the average error rate in the S1 is 2.07\% and it is 1.91\% in the S2.

\subsection{Experiment Results Analysis}
\vspace{\sectiondistance}
\label{experimentanalysis}
In this study, we through real attacks to verify the availability of \papertitleabbr\ attack on stealing the SGX-private data combined the Foreshadow attack. Besides, our experiment results demonstrate that the vulnerable PMU events and their trigger instructions are numerous, which should be noticed when designing the PMU module of the processor. Moreover, the throughput and error rate of the \papertitleabbr\ attack are measured. Even the error rate is not fully evaluated for the PMU events whose the number of trigger instructions in every scenario is larger than 200, it does not influence the conclusion that there are a lot of ways to implement the \papertitleabbr\ attack.

We consider two scenarios of \texttt{ins1} and \texttt{ins2} (shown in the Section \ref{iteartion}), this may reduce our experiment results about the number of trigger instructions for PMU events because we do not test all the combinations of \texttt{ins1} and \texttt{ins2}, but the PMU events that we obtain will be similar to the situation that we test all the combinations of \texttt{ins1} and \texttt{ins2} because that all the valid instructions are tested for every PMU event. Besides, finding part of the trigger instructions for a PMU event is enough and acceptable in this study as our most important aim is to find the vulnerable PMU events. Moreover, for every trigger instruction, we may directly construct a new trigger instruction by replacing its \texttt{nop} instruction with any of valid instructions except for the instruction that equals to the efficient instruction in the trigger instruction. This is because that the efficient instruction is enough to trigger the PMU events, the \texttt{nop} instruction is just to be an assistant instruction. We have partly verified this judgment through real experiments. Therefore, only considering the two scenarios is acceptable in this study.

\section{Mitigation to \papertitleabbr\ Attack}
\vspace{\sectiondistance}
In this section, we present some possible solutions to mitigate \papertitleabbr\ attack from both hardware and software sides. Compared to the hardware-based solutions, the software-based solutions are easier-to-deploy but may take more performance overhead, the hardware-based solutions can take less performance overhead but are not very easy to be deployed on the existing devices. Since the \papertitleabbr\ attack is a hardware vulnerability, it will be more efficient to mitigate it with the hardware-based solutions.

\subsection{Hardware-based Countermeasures}
\vspace{\sectiondistance}
\papertitleabbr\ attack is caused by the PMU module of processor, redesigning the PMU module is a fundamental solution to address it. What should be mentioned is that it is not a good choice to disable the PMU module because that it is a very significant hardware module for the software developers to optimize their applications.

The root cause of \papertitleabbr\ attack is that some PMU events triggered during the transient executions will be recorded by the corresponding PMU counters. Therefore, we propose to mitigate the \papertitleabbr\ attack by disabling the vulnerable PMU counters that record the transient execution events. This may prevent the researchers from reverse engineering the micro-architecture designs of block-box processors, but has negligible effects on the clients. Actually, reverse engineering the block-box processors is not one of the aims of PMU. In the transitional designs of PMU, if the out-of-order or speculative executions are successfully committed, the changed PMU counters will not be roll-backed, which can improve the PMU's performance on recording the events. Therefore, the drawback of this design is that the performance of the PMU module may be reduced.

Enabling PMU to record the events that are activated by the out-of-order or speculative executions but roll-backing them if those executions are not successfully committed is an improved implementation of the proposed method. In this design, the changed PMU counters will be kept if the executions are successfully committed, which can take a small performance overhead but still is able to address the \papertitleabbr\ attack. Like the current hardware modules that will be roll-backed when needed, this design can be implemented by renaming the PMU counters and putting the monitored results into the reorder buffer. The actual PMU counters only will be updated when the instructions successfully retire.

\subsection{Software-based Countermeasures}
\vspace{\sectiondistance}
In the software side, we can mitigate \papertitleabbr\ attack by disabling the PMU through microcode updates. However, similar to the method of removing the hardware PMU module that is described in the aforementioned hardware-based countermeasures, this will be bad for the software developers. Because the attack targets of the \papertitleabbr\ attack are the procedures and data owned by TEE, ensuring that the TEE is not vulnerable to \papertitleabbr\ attack is a good measurement (removing PMU from the TCB of TEE). For example, the processor vendors can make a microcode update to provide an interface for the clients to disable PMU through the Basic Input Output System (BIOS). If the clients need to use TEE, it should first disable the PMU. Otherwise, the PMU can still be enabled. At the launch step of TEE, the verification codes check whether PMU is disabled (whether the PMU's statuses violate the TCB of TEE). If PMU is enabled, it will stop to launch the TEE-private procedures and data. If not, the TEE will sequentially perform following procedures. This software-based countermeasure has been utilized by the Intel to mitigate the Dynamic Voltage and Frequency scaling (DVFS)-based vulnerabilities~\cite{qiu2019voltjockeySGX}.

\section{Discussion and Future Work}
\vspace{\sectiondistance}
In this section, we discuss the limitation and possible extensions of the proposed \papertitleabbr\ attack.

\subsection{Limitation}
\vspace{\sectiondistance}
Although we have verified the effectiveness of \papertitleabbr\ attack through real experiments, there are some limitations that should be discussed.

\subsubsection{Transient Execution Vulnerabilities}
\vspace{\sectiondistance}
\papertitleabbr\ attack is a kind of side channel attack, in which the PMU events that are recorded by the hardware PMU module in transient executions are the side channel signals. Like most of the traditional side channel attacks, it depends on other vulnerabilities (transient execution vulnerabilities in this study) to achieve the information leakage. Therefore, if the transient execution vulnerabilities are addressed, the \papertitleabbr\ attack may not become a real threat to the computer system. Fortunately, we cannot ensure that our processors are fully resistant to the transient execution attacks. Even a set of transient execution vulnerabilities are reported and mitigated, they may be just the tip of the iceberg of the processor's transient execution vulnerabilities. The undisclosed transient execution vulnerabilities can utilize the \papertitleabbr\ attack to implement the information leakage especially on the processors whose cache memory are resistant to the side channel attacks.

\subsubsection{Throughput}
\begin{figure}[htb]
\begin{center}
\subfigure[The error rate (\%) in the S1.]{ \label{fig:s1errorround}
\includegraphics[width=0.91\columnwidth]{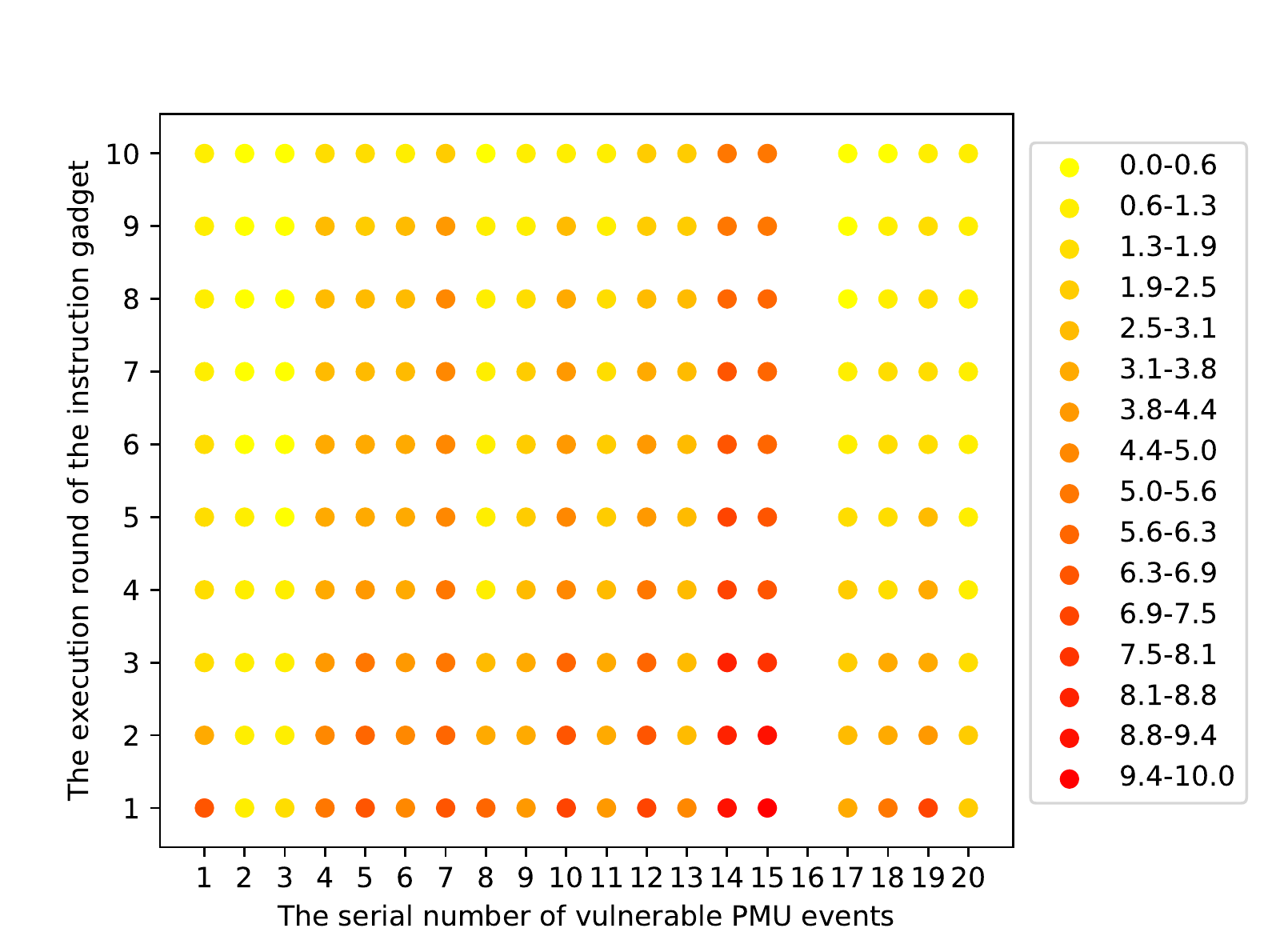}}
\subfigure[The error rate (\%) in the S2.]{ \label{fig:s2errorround}
\includegraphics[width=0.91\columnwidth]{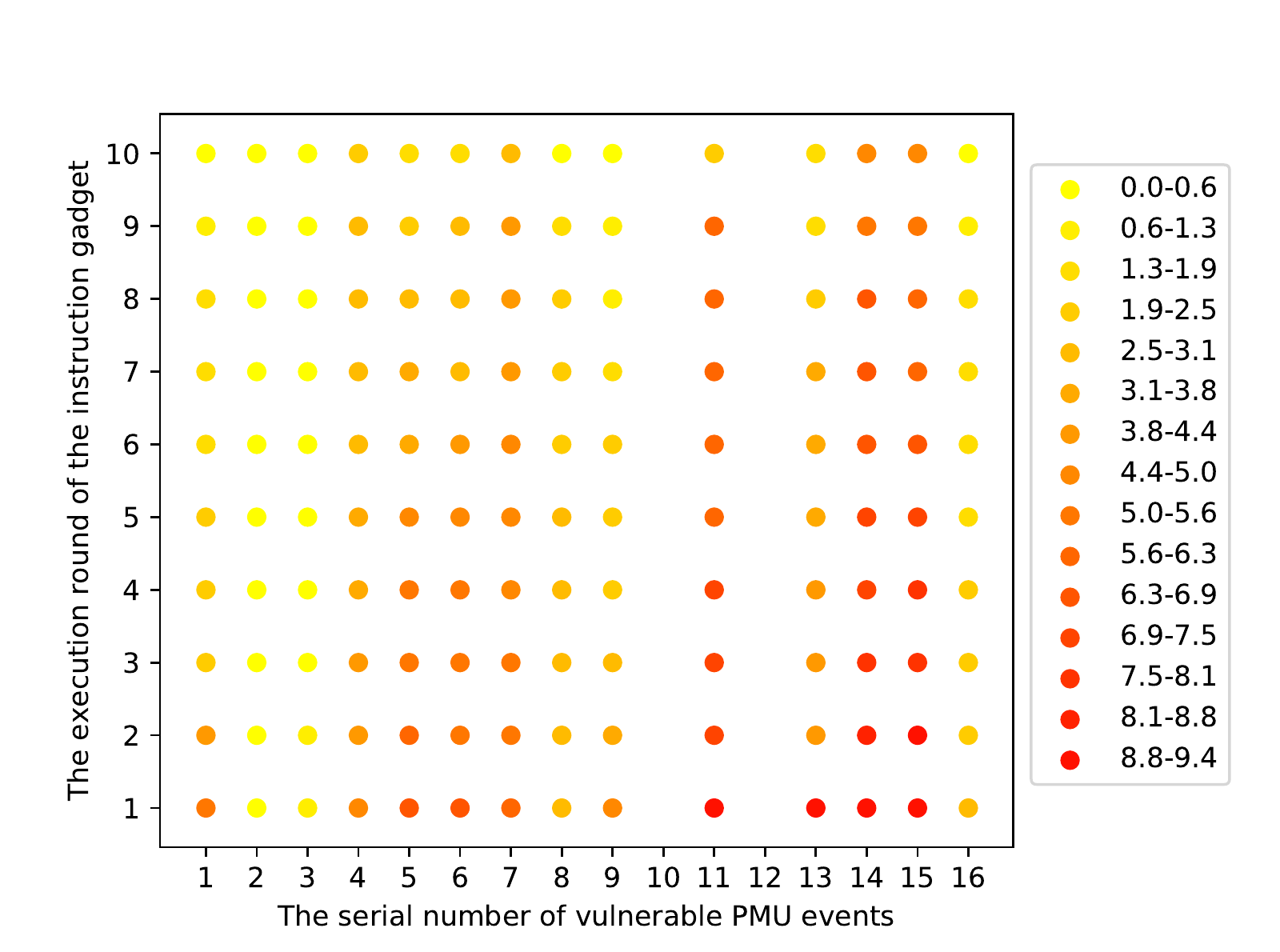}}
\caption{The error rate with different execution rounds of the instruction gadget for recovering one byte.}
\label{fig:errorround}
\end{center}
\end{figure}

\vspace{\sectiondistance}
The throughput of \papertitleabbr\ attack is lower than some of the traditional side channel attacks especially the cache side channel attacks, for example, the throughput of Foreshadow attack that depends on the Flush+Reload cache side channel attack is 27693Bps. This is because that reading the PMU counters are more time-expensive than accessing the cache memory. Besides, for every byte of the secret data, the instruction gadget is executed 10 rounds to reduce the error rate, which results in that the core instruction gadget will be executed $10*256*6=15360$ times for every byte. However, as shown in our real experiments, the \papertitleabbr\ attack can successfully recover the secret data of the Intel SGX combined the Foreshadow attack and the throughput is acceptable. Of course, we can improve the throughput by reducing the number of executions of the instruction gadget for recovering one byte of the secret data, but this will influence the error rate. Table \ref{table:throughputrounds} and Figure \ref{fig:errorround} shows our experiment results about the throughput and error rate with different execution rounds of the instruction gadget when recovering once byte, respectively. Reducing the execution round will increase the throughput but also increase the error rate and vice versa.

\begin{table}[htb]
\centering
\caption{The throughput with different execution rounds of the instruction gadget for recovering one byte.}
\label{table:throughputrounds}
\scriptsize
\begin{tabular}{c|cccccccccc}
Rounds & 1 & 2 &3 & 4 & 5 & 6 & 7 & 8 & 9 & 10\\
\hline
Throug.   & \multirow{2}{*}{5.5} & \multirow{2}{*}{2.7} & \multirow{2}{*}{1.8} & \multirow{2}{*}{1,4} & \multirow{2}{*}{1.1} & \multirow{2}{*}{0.93} & \multirow{2}{*}{0.78} & \multirow{2}{*}{0.66} & \multirow{2}{*}{0.61} & \multirow{2}{*}{0.58} \\
(KBps) &  &  & &  &  &  &  &  &  & \\
\end{tabular}
\end{table}

\subsubsection{Exception Suppression}
\vspace{\sectiondistance}
In the Listing \ref{Instructiongadget} and our aforementioned experiments, we utilize the Intel TSX to suppress the exception triggered by the secret data load instructions as it has no additional influence on the PMU events. However, not all of the processors enable Intel TSX. Using the software-based exception handling interface provided by the OS is an alternative exception suppression method. However, this will introduce some additional instructions that may also trigger the monitored PMU events, which may increase the error rate. Luckily, the exception triggered by the secret data load instructions is fixed and the exception handling instructions are the same. Therefore, the PMU events triggered by the exception handling instructions can be treated as an invariant. We measure the average error rate of the \papertitleabbr\ attack for each PMU event when using the OS-provided interface to suppress the exceptions. The experiment results are shown in the Table \ref{table:results}. The average error rate in the two scenarios when handling the exception with the software is 5.35\%, which is about 2.69X than that when using the Intel TSX to suppress the exception. The experiment result indicates that using the software to handle the exception is actually an alternative method to achieve the \papertitleabbr\ attack.

\subsection{Future Work}
\vspace{\sectiondistance}
In this study, we utilize \papertitleabbr\ attack to speculate the secret data stored in the Intel SGX by achieving the Foreshadow attack, more studies may be conducted based on the \papertitleabbr\ attack. 

\subsubsection{Achieving other Transient Execution Attacks} 
\vspace{\sectiondistance}
The Foreshadow attack is utilized as a demonstration to verify the availability of \papertitleabbr\ attack in leaking the secret data of SGX. However, the applications of the \papertitleabbr\ attack may not be limited to implement the existing transient execution attacks. The future transient execution attacks that are able to break the TEE may also utilize \papertitleabbr\ attack to complete the information leakage.

\subsubsection{Leaking other Information} 
\vspace{\sectiondistance}
The experiments utilize \papertitleabbr\ attack to leak the data accessed during transient execution attacks by judging which execution path of the instruction gadget is taken. In the future work, we can capture the execution processes of the victim procedures by measuring the PMU events happened during the executions and therefore to rebuild their execution flow. Based on which, we can speculate the functions of the victim procedures or the secret data of the procedures such as the key of AES, RSA, ECC, et al.

\subsubsection{Detecting the Transient Execution Attacks} 
\vspace{\sectiondistance}
The transient execution attacks cannot leak the secret data directly as the secret data is accessed transiently. They usually place the secret data-related information into the layouts of the processor's micro-architecture modules (cache is the most utilized module) and recover the secret data by detecting whether the layouts of the modules are changed (e.g. which element of the probe array is cached). This step may activate some special events. We can detect the transient execution attacks by judging whether the special events-corresponding counters are increased after the transient executions. 

\subsubsection{Attacking other Processors} 
\vspace{\sectiondistance}
The processor's architecture of the experiment device is Skylake in this study. However, PMU is widely enabled by most of the current processors including those by Intel, ARM, and AMD. Therefore, we can extend the \papertitleabbr\ attack to the processors that allow PMU to record the events happened during transient executions. In this extension, the attack target needs to be the data stored in the TEE of the victim processor such as Intel SGX, ARM TrustZone, or AMD Secure Encrypted Virtualization (SEV).

\subsubsection{Feeding More Instructions to \texttt{ins1} and \texttt{ins2}} 
\vspace{\sectiondistance}
In our experiments, we only replace the \texttt{ins1} and \texttt{ins2} of the instruction gadget with a single instruction, respectively. This is enough to find out the vulnerable PMU events and their part of trigger instructions. However, the \texttt{ins1} and \texttt{ins2} are not limited into a single instruction, they can be fed with multiple instructions. By doing this, we can achieve more goals. Firstly, we may get more trigger instructions. Secondly, the increment of the PMU event also can be increased, especially when the trigger instruction is repeated several times in the \texttt{ins1} or \texttt{ins2}. Finally, we may find more vulnerable PMU events as some PMU events may relay on a instruction sequence instead of a single instruction. However, this will take an explosion of the search space.

\section{Conclusion}
\vspace{\sectiondistance}
PMU is an important hardware module for software developers to profile the performance of their applications. In this study, we find that some events triggered by the instructions that do not successfully retire will also be recorded by the PMU counters, which will take unexpected data leakage. Based on this finding, we propose the \papertitleabbr\ attack, a kind of side channel attack that utilizes PMU events as side channel signals to recover the secret data that are accessed during transient executions. In order to achieve the \papertitleabbr\ attack, we design an instruction gadget to encode the secret data into its execution path that can be identified by the PMU counters. We verify the usability of the \papertitleabbr\ attack by implementing the Foreshadow attack to steal the secret data stored in the Intel SGX. Our experiments suggest that several PMU events are vulnerable. We also provide some possible hardware and software-based solutions for addressing the \papertitleabbr\ attack. At the end of the paper, we analyze the limitation and possible extensions of this study. 


\bibliographystyle{IEEEtranS}
\bibliography{refs}

\end{document}